 \RequirePackage[displaymath]{lineno}
\documentclass[pre,superscriptaddress,twocolumn,showpacs,preprintnumbers,amsmath
amssymb]{revtex4}
\usepackage[pdftitle={Prepare for PRE},bookmarks=true,citecolor=red,colorlinks=false]{hyperref}
\usepackage{amsmath}
\usepackage{graphicx}
\usepackage{dcolumn}
\usepackage{bm}
\usepackage{amssymb}
\usepackage{amsthm}
\usepackage{wasysym}
\usepackage{natbib}
\usepackage{xcolor}

\usepackage{CJK}
\newcommand{\upd}{\mathrm{\,d}}
\newcommand{\figref}[1]{~\ref{#1}}
\newcommand{\red}[1]{\textcolor{black}{#1}}
\newcommand{\blue}[1]{\textcolor{blue}{#1}}
 \switchlinenumbers

\begin{document}

  \linenumbers
\preprint{APS/PRE}
\begin{CJK*}{GB}{gbsn} 
\title{Second order structure  function  in fully developed turbulence}

\author{Y.X. Huang (»ÆÓÀÏé)\footnote{Present address: Environmental Hydroacoustics
Lab,
 Universit\'e Libre de
Bruxelles,
av. F.D. Roosevelt 50 - CP 194/05, B-1050 Brussels} }
\email{yongxianghuang@gmail.com}
\affiliation{
        {Shanghai Institute of Applied
         Mathematics and Mechanics,
        Shanghai University, 200072 Shanghai,  China}
}
\affiliation{{Univ Lille Nord de France, F-59000 Lille, France}}
\affiliation{{USTL, LOG, F-62930 Wimereux, France}}
\affiliation{{CNRS, UMR 8187, F-62930 Wimereux, France}}

\author{F. G. Schmitt}%
 \email{francois.schmitt@univ-lille1.fr}
\affiliation{{Univ Lille Nord de France, F-59000 Lille, France}}
\affiliation{{USTL, LOG, F-62930 Wimereux, France}}
\affiliation{{CNRS, UMR 8187, F-62930 Wimereux, France}}
 \author{Z.M. Lu (¬־Ã÷)}
 \affiliation{
        {Shanghai Institute of Applied
         Mathematics and Mechanics,
        Shanghai University, 200072 Shanghai,  China}
}
\author{P. Fougairolles}
\affiliation{
        {CEA, DTN/SE2T/JIEX, 38054 Grenoble, France}}
\affiliation{
        {LEGI, CNRS/UJF/INPG, UMR 5519, 38041 Grenoble, France}}
\author{Y. Gagne}
\affiliation{
        {LEGI, CNRS/UJF/INPG, UMR 5519, 38041 Grenoble, France}}
\author{Y.L. Liu (ÁõÓî½)}
\affiliation{
        {Shanghai Institute of Applied
         Mathematics and Mechanics,
        Shanghai University, 200072 Shanghai,  China}
}

\date{\today}

\begin{abstract}
  We relate the second
order structure function of a time series with the power spectrum of the
original variable, taking  an assumption of statistical stationarity. With
this approach, we find that the structure function is strongly influenced by
the large scales. The large scale contribution and the contribution range are
respectively 79\% and 1.4 decades for a  Kolmogorov -5/3  power spectrum. We
show numerically that a single scale influence range,  
over smaller scales is
about 2 decades. We argue that the structure function  is
not a good method to extract the scaling exponents when the data possess
large energetic scales. An alternative methodology, the arbitrary order Hilbert
 spectral analysis which may constrain this influence within 0.3 decade, is
proposed to characterize the {scaling} property
directly in an amplitude-frequency space. An analysis of passive scalar
(temperature)
 turbulence time series is presented to show the influence of large scale
structures in real turbulence, and the efficiency of the Hilbert-based
methodology. The corresponding scaling exponents $\zeta_{\theta}(q)$ provided by
the Hilbert-based approach indicate that the passive scalar turbulence field may
be less intermittent than what was previously believed.

\end{abstract}

\maketitle
\end{CJK*}

\section{Introduction}

The most intriguing property  of fully developed turbulence is its scale
invariance, characterized by a sequence of scaling exponents
\cite{Kolmogorov1941,Frisch1995}. Since Kolmogorov's 1941 milestone work, 
structure function analysis is  widely used  to extract these scaling exponents \cite{Anselmet1984,Lepore2009,Sreenivasan1997,Lohse2010}. The
second order structure function is written as
(we work in temporal space here, through Taylor's hypothesis)

\begin{equation}
  S_2(\ell)=\langle \Delta u_{\ell}(t)^2 \rangle \sim \ell^{\zeta(2)}
\end{equation}
where $\Delta u_{\ell}(t)=u(t+\ell)-u(t)$ is the velocity increment, $\ell$ is separation time,
and according to K41 theory $\zeta(2)=2/3$ in
the inertial range \cite{Kolmogorov1941,Frisch1995}.
However, the structure function itself is seldom investigated in detail
\cite{Bacry1993,Nicholspagel2008}.
Structure functions have been considered as `poor man's wavelets' by some authors \cite{Bacry1993}.
This was mainly linked to a bound in the singularity range that can be grasped by structure functions.

In this paper, we address another issue, the contribution from the large
scale structures part and the influence
range of a single scale. By taking a statistic stationary assumption
 and the Wiener-Khinchin theorem \cite{Percival1993}, we  relate the second
order structure function to the Fourier power spectrum of the original
 velocity
\cite{Frisch1995}.  We define a cumulative function $\mathcal{P}(f,\ell)$
to characterize the relative contribution
of large scale structures, where $\ell$ is the separation scale.
 It is found
that for a pure Kolmogorov 5/3 spectrum the  large scale contribution range
is more than 1.4 decades and the corresponding relative contribution  is
about 79\%.   We show an analysis of experimental homogeneous and nearly
isotropic turbulent velocity data base. The compensated spectra provided by
different methods show that, due to the influence of large scale structures, the
second order structure function predicts a shorter inertial range than other
approaches. The cumulative function estimated from the turbulence database shows
that the largest contribution of the second order structure function is coming
from the large scale part.
We then check the influence of a single scale by using fractional Brownian
motion (fBm) simulations. We show that the influence range  over smaller
scales
is as large as two decades.
 We also show that the Hilbert-based methodology
\cite{Huang2008EPL,Huang2008TSI,Huang2009PHD}
could constrain this effect within 0.3 decade. We finally analyze a
passive scalar (temperature) time series, 
in which the large scale `ramp-cliff' structures
play an important role \cite{Sreenivasan1991,Shraiman2000,Warhaft2000}.
Due to the presence of strong ramp-cliff structures,
the structure function analysis fails. However, the Hilbert-based approach
displays a clear inertial range. The corresponding scaling exponents
are quite close to the scaling exponents of longitudinal velocity, indicating
a less intermittent passive scalar statistics than what was believed before.

\red{
This paper is organized as follows. In section \ref{sec:HHT}, we
  briefly 
introduce the empirical mode decomposition and  arbitrary
order Hilbert spectral analysis.  By considering Wiener-Khinchin theorem,
an analytical model for the second order structure
function is proposed in section \ref{sec:SF}. In section \ref{sec:PST}, analysis
results of passive turbulence (temperature) experiment data are presented.
We draw the main results and conclusions in section \ref{sec:CL}.}

\section{Arbitrary order Hilbert spectral analysis}\label{sec:HHT}
\red{\blue{Arbitrary order Hilbert spectral analysis is an extended version of the
 Hilbert-Huang transform (HHT)} \cite{Huang1998,Huang1999}. It is
designed to characterize scale invariant properties directly in an amplitude-frequency
space \cite{Huang2008EPL,Huang2008TSI,Huang2009PHD}. The method possesses
two steps: Empirical Mode Decomposition (EMD) and Hilbert spectral analysis.
We present a briefly introduction below.}

\subsection{Empirical Mode Decomposition}

\begin{figure}[htb]
\centering
 \includegraphics[width=0.95\linewidth]{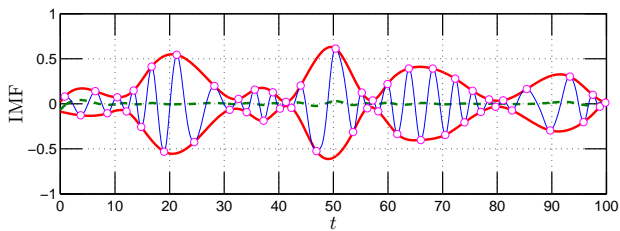}
  \caption{\red{(Color online) An example of IMF from EMD: local extrema points ($\ocircle$),
  envelope (thick solid line) and running mean (dashed line). It indicates both
amplitude and
  frequency modulations of the Hilbert-based method.} }\label{fig:IMF}
\end{figure}

\red{The idea behind  EMD is to consider the multi-scale properties of
real
time series.
Then Intrinsic Mode Functions (IMFs) are proposed as
 mono-scale components. To be an IMF, a function has to
satisfy the following two conditions:  (\romannumeral1)
 the difference between
the number of local extrema and the number of zero-crossings must be zero or one; (\romannumeral2) the running mean value
of the envelope
defined by the local maxima and the envelope defined by the local minima is zero \cite{Huang1998,Huang1999}. 
Figure \ref{fig:IMF} shows an example
of IMF from EMD, showing both amplitude- and frequency-modulations of Hilbert-based
method \cite{Huang1998,Huang2009PHD}.
The EMD algorithm, a sifting process,  is then designed to
decompose a given time series $x(t)$ into a sum of IMF modes $C_i(t)$}

\red{\begin{equation}
x(t)=\sum_{i=1}^{n} C_i(t)+r_n(t)
\end{equation}
where $r_n(t)$ is the residual, which is either a constant or a monotone
function \cite{Huang1998,Huang1999,Rilling2003}.  Unlike classical decompositions
(Fourier, Wavelet, etc.),  there is no basis assumption before the
decomposition.  In
other words, the basis
is deduced by the data themselves, which means \red{that this}  is a completely
data-driven method with very local \red{abilities} in \red{the} physical domain
\cite{Huang1998,Flandrin2004}.}

\subsection{Arbitrary order Hilbert Spectral Analysis}
\red{After obtaining the IMF modes,  Hilbert transform
\cite{Cohen1995,Flandrin1998} is
 applied to each IMF}
 
 \red{
 \begin{equation}
\overline{C}_i(t)=\frac{1}{\pi}P\int \frac{C_i(t')}{t-t'}\upd t'
\end{equation}
where $C_i(t)$ is the $i$th IMF mode and $P$ indicates Cauchy principal value.
Then the analytical signal is constructed
$C^{A}_i(t)=C_{i}(t)+j\overline{C}_i(t)$.
 The instantaneous frequency $\omega$ and amplitude $\mathcal{A}$ are
estimated by}

\red{
\begin{equation}
\omega(t)=\blue{\frac{1}{2\pi}}\frac{\upd \theta}{\upd t},\quad \mathcal{A}=\left(
C_i^2(t)+\overline{C}_i^2(t) \right)^{1/2}
\end{equation}
in which $\theta=\arctan \overline{C}_i(t)/C_i(t)$. Since the Hilbert transform
is a singularity integration, the $\omega$ thus \red{have} very local ability in
spectral space and \red{are} free with limitation of the  Heisenberg-Gabor uncertainty
principle \cite{Huang2005a,Cohen1995,Flandrin1998}. \red{ After performing this
on all modes series obtained from the analyzed series $x(t)$, one obtains
 a} joint pdf
$p(\omega,\mathcal{A})$, which  can be extracted from $\omega$ and $\mathcal{A}$
\cite{Long1995,Huang2008EPL,Huang2008TSI,Huang2009PHD}.
The arbitrary order Hilbert marginal spectrum is defined by considering a
marginal integration  of the joint pdf $p(\omega,\mathcal{A})$, which reads as}

\red{
\begin{equation}
\mathcal{L}_q(\omega)=\int p(\omega,\mathcal{A}) \mathcal{A}^q \upd \mathcal{A}
\end{equation}
where $q\ge0$, $\omega$ is the instantaneous frequency, $\mathcal{A}$ the
amplitude \cite{Huang2008EPL,Huang2008TSI,Huang2009PHD}.
In case of scale invariance, we expect}

\red{
\begin{equation}
\mathcal{L}_q(\omega)\sim \omega^{-\xi(q)}
\end{equation}
  We have  shown \red{elsewhere that} $\xi(q)=1+qH$ for fractional Brownian
 motion, where $H$ is Hurst number
\cite{Huang2008EPL,Huang2008TSI,Huang2009PHD}.
This generalized Hilbert spectral analysis has been successfully applied
to turbulence velocity \cite{Huang2008EPL},
daily river flow discharge \cite{Huang2009Hydrol}, surf zone
\cite{Schmitt2009JM}, etc., to characterize the scale invariance
directly in the amplitude-frequency space \cite{Huang2009PHD}.}

\red{ The main
drawback of the Hilbert-based methodology is its first step, Empirical Mode
Decomposition, which is an algorithm in practice without rigorous  mathematical
foundation \cite{Huang1998,Huang2005a}.  Flandrin and his co-workers have
obtained some theoretical results
on the EMD method \cite{Flandrin2004,Rilling2006,Rilling2008,Rilling2009}.
However, more theoretical work is still needed to fully mathematically understand
this method.}

\section{second order structure function}\label{sec:SF}

\begin{figure}[!htb]
\centering
 \includegraphics[width=0.95\linewidth]{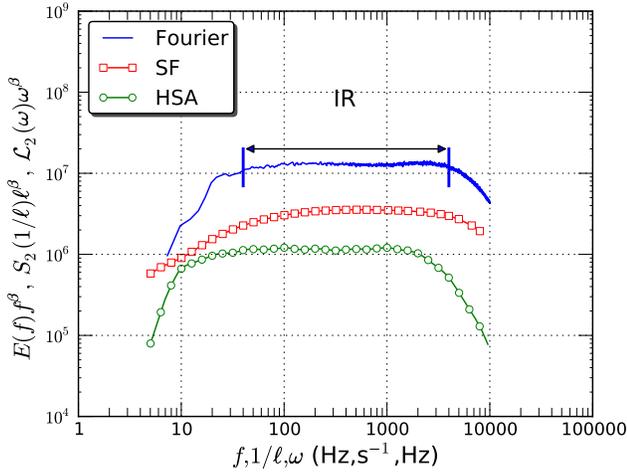}
  \caption{(Color online) Compensated spectra  of  transverse  velocity. A plateau is observed
on the range  $40<f<4000$ Hz for Fourier spectrum (solid line) and  $20<f<2000$
Hz  for Hilbert spectrum ($\ocircle$), respectively. For comparison, the
compensated spectra for the second order structure function ($\square$) is also shown. The compensated
values $\beta$ are estimated case by case. For display convenience, the curves
have been vertically shifted.}\label{fig:compspv}
\end{figure}

 \begin{table}[!h]
 \begin{center}
 \begin{tabular}{|c|c|c|c|c|c|c|c|c|}
 \hline
$f$ (Hz) & 0.01 & 0.04& 0.1 & 0.2 &0.5& 1 & 10& 100 \\
 \hline
 $\mathcal{P}$ (\%) & 0.46  &2.95 & 9.91 & 24.0& 62.7 & 78.6 & 95.3 & 99.0\\
 \hline
 $\mathcal{Q}$ (\%) & -2.3  & -14.1 & -44.4 & -83.5 & 1.8 &49.0 &88.5 & 97.6\\
  \hline
 \end{tabular}
 \end{center}
 \caption{\red{Index values of analytical expressions 
$\mathcal{P}(f,1)$ and $\mathcal{Q}(f,1)$ with $\beta=5/3$ for several frequencies.}}
 \label{tb:analyticalPQ}
 \end{table}

\begin{figure}[!htb]
\centering
 \includegraphics[width=0.95\linewidth]{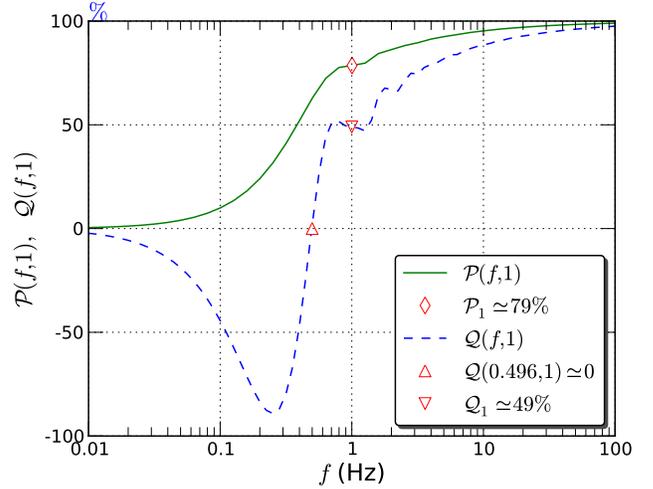}
  \caption{(Color online) \red{Semilog plot of analytical expressions of $\mathcal{P}(f,1)$ (solid
line) and $\mathcal{Q}(f,1)$  (dashed line) with
$\beta=5/3$ on the range $0.01<f<100$ \blue{Hz}. Symbols are respectively the index
values of 
$\mathcal{P}_1(1)\simeq79\%$ ($\diamondsuit$), the large scales part contribution to the second order structure function,
$\mathcal{Q}(0.496,1)\simeq0$ ($\triangle$), the zero-crossing point of the autocorrelation function, and
$\mathcal{Q}_1(1)\simeq49\%$ ($\triangledown$), the large scales part contribution to the autocorrelation function, see also Table \ref{tb:analyticalPQ}.}
}\label{fig:analytical}
\end{figure}

\begin{figure}[!htb]
\centering
 \includegraphics[width=0.95\linewidth]{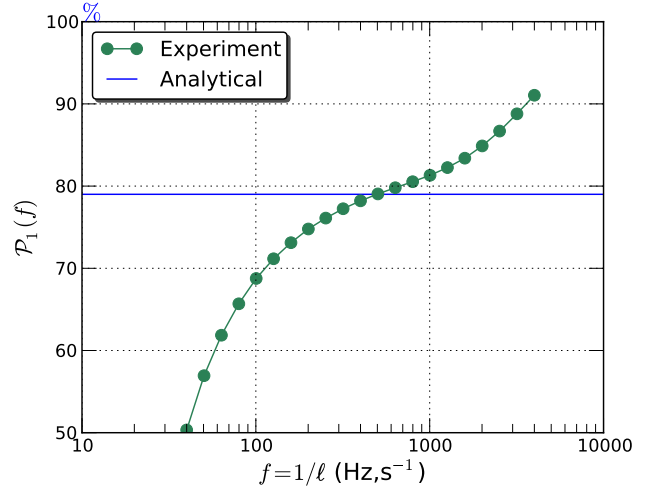}
  \caption{(Color online) Cumulative function $\mathcal{P}_1(f)$ estimated from turbulent
  experimental data for transverse velocity on the inertial range
$40<f<4000\,$Hz.
The analytical expression \red{for} $\mathcal{P}$ shows
 $\mathcal{P}_1\simeq 79 \%$ (horizontal solid line).
 \red{We note}
that all $\mathcal{P}_1\ge 50\%$, which means that  most contribution of the
second order structure function comes from the large scales part $f<1/\ell$.
}\label{fig:Q1}
\end{figure}

\begin{figure}[!htb]
\centering
 \includegraphics[width=0.95\linewidth]{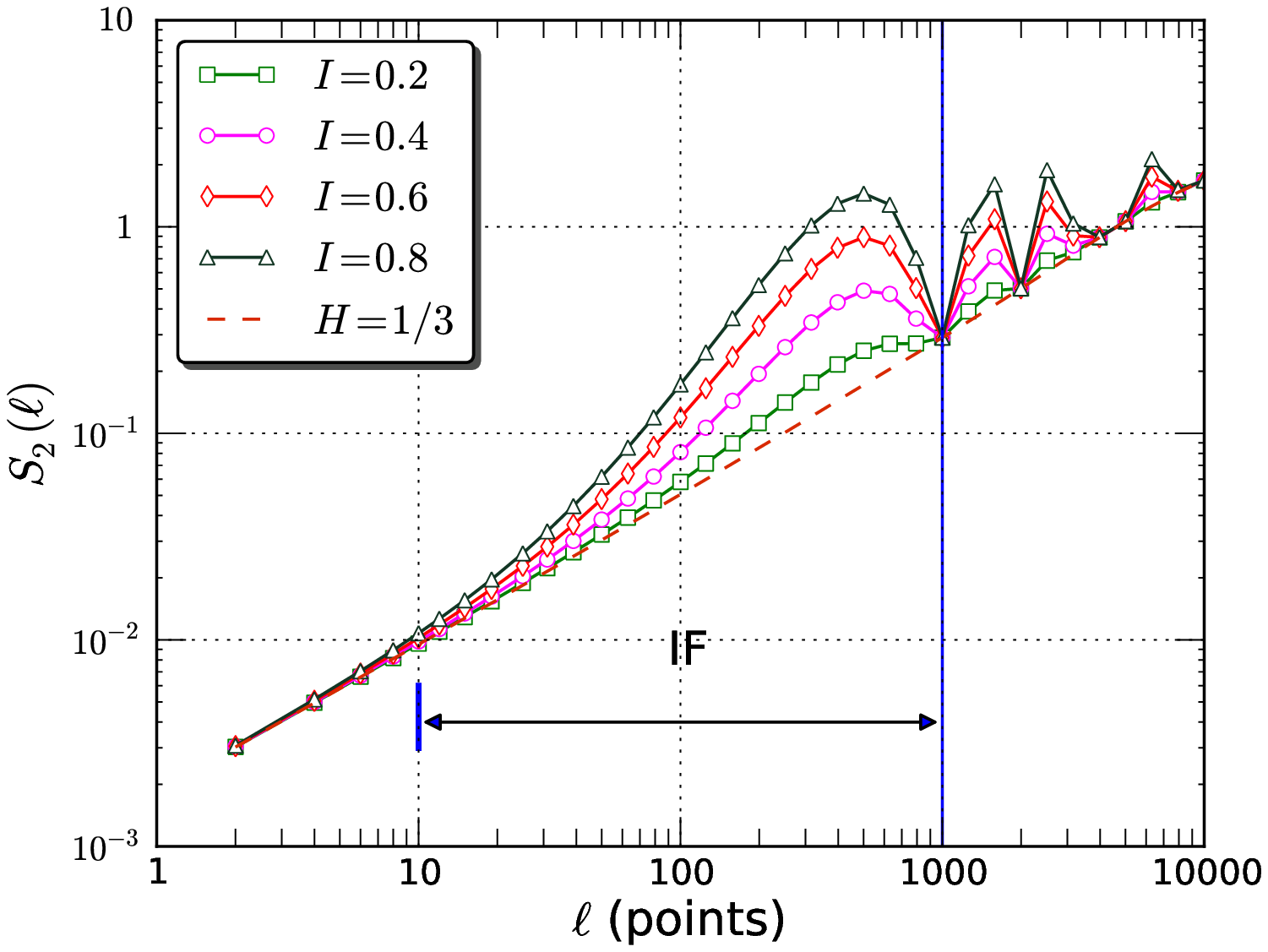}
  \caption{(Color online) Periodic effect on the second order structure function with various
intensities $I$, where the vertical line illustrates the location of the
perturbation sine wave. }\label{fig:fbmsf}
\end{figure}

\begin{figure}[!htb]
\centering
 \includegraphics[width=0.95\linewidth]{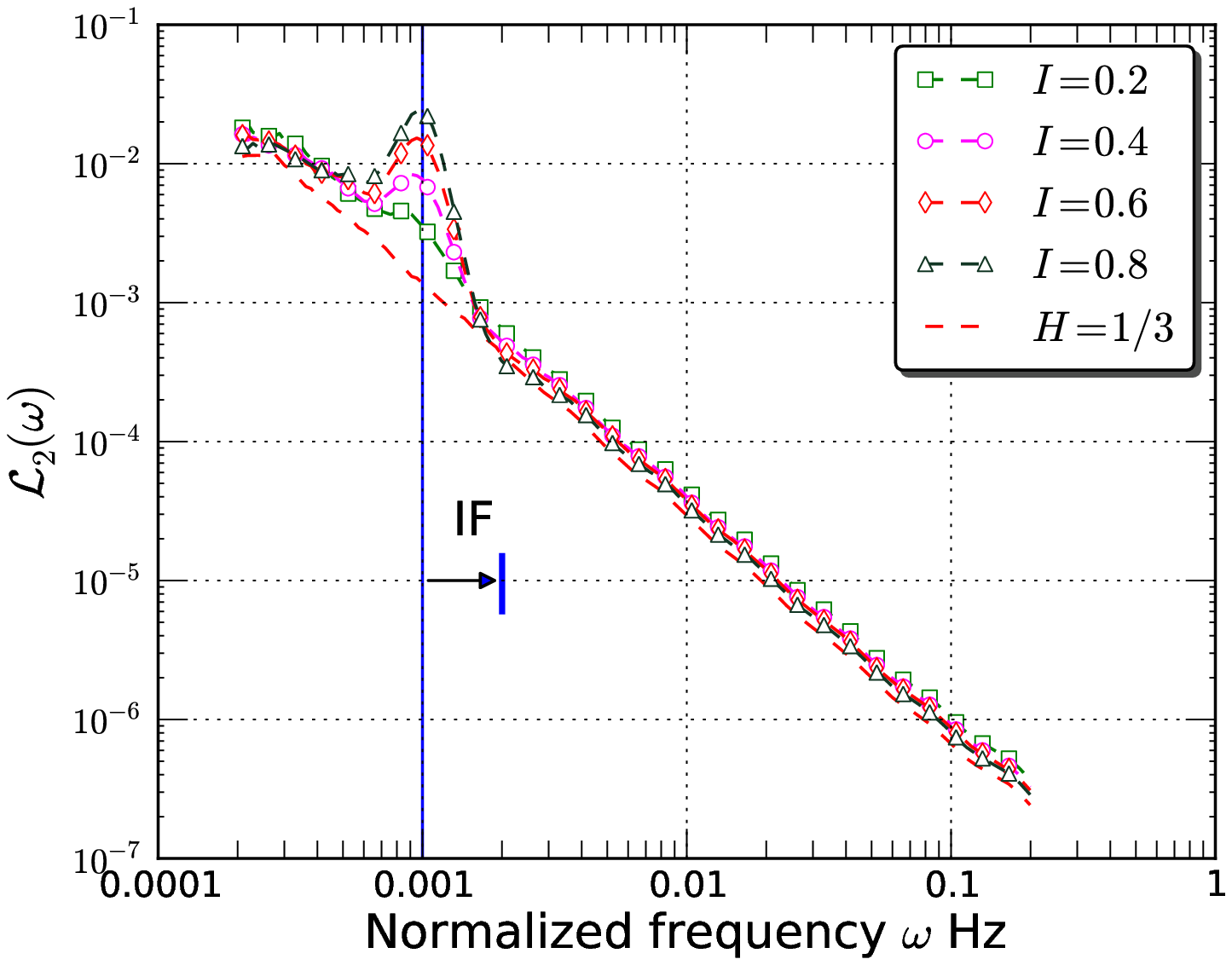}
  \caption{(Color online) Periodic effect on the second order Hilbert marginal spectrum with
various intensities $I$, where the vertical line illustrates the location of the
perturbation sine wave. }\label{fig:fbmhsa}
\end{figure}

\red{The structure function is the most widely used method in turbulence
research to extract the
scaling exponents \cite{Monin1971,Anselmet1984,Frisch1995,
Sreenivasan1997,Lepore2009,Lohse2010}. It has also been used in may other
fields to characterize the scale invariance properties of time series, e.g.
climate data
\cite{Schmitt1995GRL}, financial research
\cite{Schmitt1999}, to quote a few.
The relationship between the second order structure function and the corresponding
Fourier power spectrum has been
 investigated previously by Lohse and
M{\"u}ller-Groeling \cite{Lohse1995PRL,Lohse1996PRE}.
They obtained an analytical expression of Fourier power spectrum for
turbulent velocity by considering a Batchelor fit for the second order structure
functions. They found that the energy pileups at the ends of scaling ranges in
Fourier space, which leads to a bottleneck effect in turbulence.
Here we
focus on another aspect of the second order structure function, the scale
contribution
and contribution range
from the large scale part.}

Considering a statistical stationary assumption and the Wiener-Khinchin theorem
\cite{Percival1993}, we can  relate the second order structure function to the
Fourier power spectrum of the original velocity \cite{Frisch1995}

\begin{equation}
S_2(\ell)=\langle \Delta u(\ell)^2\rangle=\int_{0}^{+\infty}E_u(f)(1-\cos(2\pi f \ell))\upd
f\label{eq:2sf}
\end{equation}
where we neglect a constant in front of the integral, and $E_u(f)$ is the
Fourier power spectrum of the velocity. Let us introduce a cumulative function

\begin{equation}
\mathcal{P}(f,\ell)=\frac{\int_{0}^{f}E_u(f')(1-\cos(2\pi f' \ell))\upd f'}
{\int_{0}^{+\infty}E_u(f')(1-\cos(2\pi f' \ell))\upd f'}\times 100\%
\label{eq:csf}
\end{equation}
$\mathcal{P}(f,\ell)$ is increasing from 0 to 1,
and measures the relative contribution to the second order structure function
from 0 to  $f$. We are particular concerned by the case $f=1/\ell$, \red{$\mathcal{P}_1(f)=\mathcal{P}(f,\ell)\vert_{f=1/\ell}$},
which measures the relative contribution from large scales. If we assume a power law for the spectrum

\begin{equation}
  E_u(f)=cf^{-\beta},\, c>0 \label{eq:pl}
\end{equation}
when substituted into Eq. \eqref{eq:2sf}, this gives a divergent
 integral for some values of $\beta$.  The convergence condition requires
$1<\beta<3$ \cite{Frisch1995}. \red{In the appendix, we
 derive an analytical expression for $S_2(\ell)$}
  
  \red{
\begin{equation}
S_2(\ell)=\frac{c\pi^{\beta-\frac{1}{2}} \,\Gamma(\frac{3}{2}-\frac{\beta}{2})}
 {(\beta-1)\,\Gamma(\frac{\beta}{2})}\ell^{\beta-1}\label{eq:analyticalSF}
\end{equation}}
\red{and for $\mathcal{P}(f,\ell)$
\begin{widetext}
\begin{equation}
 \mathcal{P}(f,\ell)= 
\frac{1}{a(\beta)}\bigg\{ (3-\beta)(\cos(f)-1)f^{1-\beta} +
g(f,\beta)\,f^{3-\beta}
\bigg\}
\times 100\%
\label{eq:analyticalP}
\end{equation}
\end{widetext}
in which
$a(\beta)=\sqrt{\pi}\,(3-\beta)\,2^{1-\beta}\,
\Gamma(3/2-\beta/2)\Gamma(\beta/2)^{-1}$, and
$g(f,\beta)={}_{1}F_2(3/2-\beta/2,3/2,5/2-\beta/2,-f^2/4)$ is  a generalized
hypergeometric function
\cite{Abramowitz1970Handbook}. }
 For fully developed turbulence, the Kolmogorov
spectrum corresponds to $\beta=5/3$
\cite{Kolmogorov1941,Frisch1995}.

We  apply here the above \red{approach} to a database from an experimental  homogeneous and nearly
isotropic turbulent channel flow at downstream $x/M=20$, where $M$ is the mesh size. The flow
is characterized by a Taylor microscale based Reynolds number $Re_{\lambda}=720$
and the sampling frequency is $f_s=40,000\,$Hz \cite{kang2003dta}.  The detail
of this
experiment can be found in Ref. \cite{kang2003dta}.
 Figure\figref{fig:compspv} shows the compensated spectra for
transverse velocity components on the range $5<f<10,000\,$Hz, in which  the
spectra
are estimated by  Fourier analysis (solid
line) \cite{kang2003dta}, the second order structure function ($\square$),
and the
arbitrary order Hilbert spectral analysis
($\ocircle$) \cite{Huang2008EPL,Huang2009PHD}, respectively. The compensated
values $\beta$ are estimated
case by case. For comparison convenience, \blue{we represent  the structure
function  as a function of $f=1/\ell$}.  Except for the structure function, there is a plateau which is more than two
decades wide. We also note that the curves provided by
second order structure function
 and the Fourier power spectrum are not identical
with each other, which is required by Eq. \eqref{eq:2sf}. This  has been reported
by several
authors  \cite{Frisch1995,Hou1998,Nelkin1994}.
The difference may come from the finite  scaling range  \cite{Nelkin1994,Hou1998}
and also violation of the  statistical stationary assumption
\cite{Huang2009PHD}.

\red{We note that $\mathcal{P}(f,\ell)$ is independent
of $\ell$ since we assume a pure power law relation \eqref{eq:pl}, see the appendix for more detail. Below we only consider
the case $\ell=1\,$\blue{s}, e.g. $\mathcal{P}(f,1)$. We concentrate on the large scales ($f<1$\,\blue{Hz}) contribution
to the second order structrue function, e.g. $\mathcal{P}_1(1)=\mathcal{P}(f,1)\vert_{f=1}$, which measures the contribution from large scales. Figure \ref{fig:analytical}  and Table \ref{tb:analyticalPQ} show respectively  the analytical curve
$\mathcal{P}(f,1)$ and various index values on
the range $0.01<f<100$\,\blue{Hz}  for a pure Kolmogorov power law by taking
$\beta=5/3$. The contribution from the large scales part ($f<1$\,\blue{Hz}) is 79\% ($\diamondsuit$), see Table \ref{tb:analyticalPQ}.}
The contribution from the first decade   large scales,
$0.1<f<1$\,\blue{Hz}, is about 69\%.  For the second decade, $0.01<f<0.1$\,\blue{Hz}, the contribution
is about 9.5\%. The large scale contribution range of the second order structure
function is more than 1.4 decades \red{ if we neglect  the 3\% contribution from $f<0.04$\,\blue{Hz},
 see Table \ref{tb:analyticalPQ}}.  \red{ We have given elsewhere an
analytical model
for the autocorrelation function of velocity increments based on the same
idea \cite{Huang2009EPL}.  It writes as}

\red{
\begin{equation}
 R(\ell,\tau)=\int_{0}^{\infty}
E_u(f)(1-\cos(2\pi f\ell))\cos(2\pi f\tau)\upd f
\end{equation}
in which $\ell$ is \red{the} separation time and $\tau$ is \red{the} time delay \cite{Huang2009EPL}.
We are particularly concerned \red{with} the case $\tau=\ell$, \blue{in which $R(\ell,\tau)$ takes its minimum value} \cite{Huang2009EPL}. Power law behavior is
found as $R(\ell,\tau)\vert_{\tau=\ell}\sim \ell^{\beta-1}$ if one
substitutes Eq. \eqref{eq:pl} into the above equation. The corresponding
cumulative function reads as}

\red{
\begin{equation}
 \mathcal{Q}(f,\ell)=\frac{\int_{0}^{f}
E_u(f') (1-\cos(2\pi f' \ell ))\cos(2\pi f' \ell )\upd f'}{\int_{0}^{\infty}
E_u(f')(1-\cos(2\pi f' \ell ))\cos(2\pi f' \ell )\upd f'}\times 100\%
\end{equation}
Again, assuming the pure power law of Eq. \eqref{eq:pl}, we have an analytical
expression \red{for the} above equation, \blue{see Eq. \eqref{eq:analyticalQ} in the Appendix.} }

\red{ For
comparison, the
analytical expression $\mathcal{Q}$ with $\beta=5/3$ is also shown as \red{a} dashed
line in Fig. \ref{fig:analytical}. 
We note that $\mathcal{Q}$
crosses zero at $f\simeq 0.496$\,\blue{Hz} ($\triangle$),  see  also Table \ref{tb:analyticalPQ},
which indicates that at this position,  contributions from large scales
$f\lesssim 0.5$\,\blue{Hz}  are vanishing (canceled by themselves). It
indicates that the large scale contribution range is about 0.3 decade, e.g. $0.496<f<1$\,\blue{Hz}}, and the
contribution itself is found to be 49\%, \red{see Table \ref{tb:analyticalPQ}}. This explains why
the \blue{ minimum value of the
autocorrelation function of the velocity increments} is a better indicator of the inertial range than
structure functions \cite{Huang2009EPL}.
The corresponding  $\mathcal{P}_1=\mathcal{P}(f,\ell)\vert_{f=1/\ell}$ based on
$E_u(f)$ from the
experimental data are shown in Fig.\figref{fig:Q1} for transverse
velocity on the range $40<f<4000\,$Hz, which is the inertial range predicted by
Fourier power spectrum, see Fig. \ref{fig:compspv}.  \red{The analytical value of $\mathcal{P}_1(1)\simeq79\%$ provided by Eq. \eqref{eq:analyticalP} is shown as a solid line. Below this line, the second order structure function is influenced by both the finite length of power law and, more importantly, large scale structures, see next paragraph. Above this line, it is thus influenced by the finite length of the power law (or viscosity).}  The index value of
$\mathcal{P}_1$ is \red{significantly}
larger than 50\%, showing that the largest contribution of the second order
structure function is coming from the large scale part.

We then consider the influence of a single scale. We simulate a
fBm time series $x(t)$ with Hurst number $H=1/3$, corresponding to the Hurst
value of turbulent velocity.
A sine wave is superposed to the  normalized  fBm data  with frequency
$f_0=0.001$\,\blue{Hz}
and various intensities $I$: $x(t)=x(t)/\textrm{Var}(x)+I \sin(2\pi f_0 t)$. We
then  perform  structure function analysis and Hilbert spectral
analysis on these data. Figure~\ref{fig:fbmsf} shows the second order structure
function. It is strongly influenced by the periodic component
\cite{Huang2008EPL}. The influence range down to the small scale is as large as
2 decades.  \red{It indicates that  the structure functions is strongly influenced by a large energetic scale  structures, e.g. coherent structures.} Figure
\ref{fig:fbmhsa} shows the corresponding second order Hilbert marginal spectrum
where the
 influence down to the small scale is constrained within 0.3 decade. It  might
be linked to the fact that the first step of the arbitrary order
Hilbert spectral analysis, the empirical mode decomposition, acts a  dyadic
filter bank for several types of time series
\cite{Huang2008EPL,Flandrin2004,Wu2004a}.

\section{Passive scalar turbulence}\label{sec:PST}

\begin{figure}[!htb]
\centering
 \includegraphics[width=0.95\linewidth]{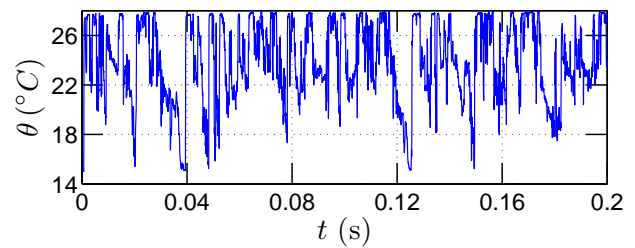}
  \caption{A 0.2s portion of the temperature time series, showing strong
ramp-cliff structures.}\label{fig:portion}
\end{figure}

\begin{figure}[!htb]
\centering
 \includegraphics[width=0.95\linewidth]{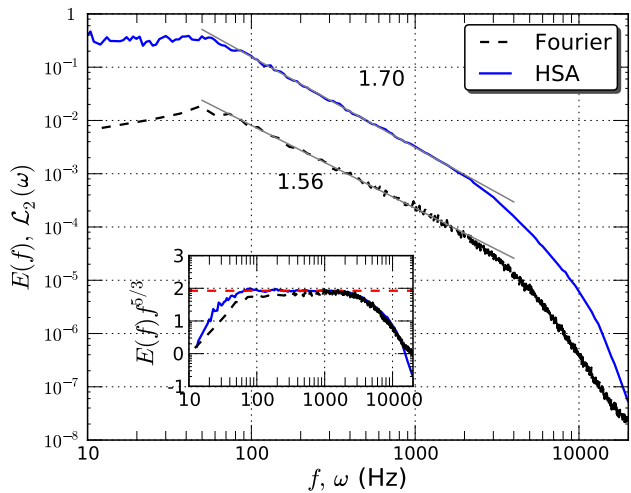}
  \caption{(Color online) \red{Fourier power spectrum and Hilbert marginal spectrum
  for temperature. Compensated spectra by $f^{5/3}$ are shown as inset. Both
methods predict power law behavior on the range
  $80<f<2000\,$Hz. For display convenience, the curves
have been vertical shifted.}}\label{fig:Tspectrum}
\end{figure}

\begin{figure}[!htb]
\centering
 \includegraphics[width=0.95\linewidth]{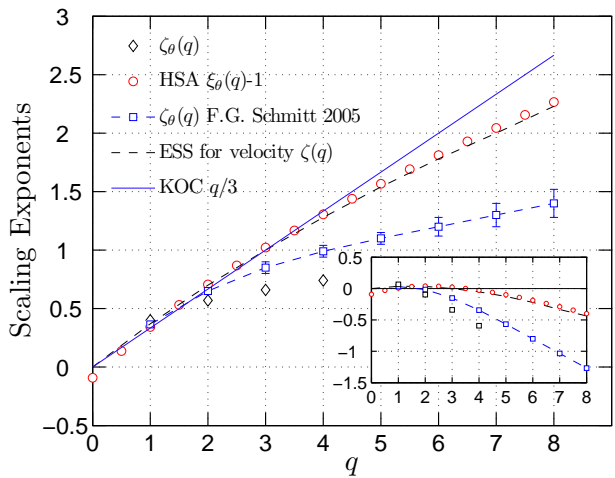}
  \caption{(Color online) Scaling exponents for passive scalar, which is estimated by
Hilbert-based approach
  $\xi_{\theta}(q)-1$ ($\ocircle$), and the structure functions
$\zeta_{\theta}(q)$ ($\diamondsuit$). For comparison, the scaling exponents
compiled by \citet{Schmitt2005} ($\square$) for passive
scalar, and  compiled by \citet{Arneodo1996} for the velocity (dashed-line) are
also shown. }\label{fig:scaling}
\end{figure}

The above arguments and results indicate that the structure functions are strongly
influenced by the large scales  and that this approach is not a good methodology to extract
the scaling exponents when the data possess    large energetic scale structures. This is the case of scalar turbulence:
ramp-cliff structures are an important signature
 of the passive scalar \cite{Sreenivasan1991,Shraiman2000,Warhaft2000,Celani2000}.
To consider this experimentally, we analyze here a temperature time series
 obtained in a shear layer  between a jet flow  and a cross flow.  The bulk Reynolds number is about $Re=60000$. The initial
temperature of the two flows are $T_{J}=27.8 ^{\circ} C$ and $T=14.8 ^{\circ} C$. The
measurement location is close to the nozzle of the jet. Figure \ref{fig:portion} shows a 0.2s portion
temperature data,   illustrating strong ramp-cliff structures.

\red{Figure \ref{fig:Tspectrum} shows the Fourier power spectrum (dashed line) and
Hilbert marginal spectrum (solid line), where the inset shows the compensated
spectra by $f^{5/3}$. Both methods predict a more than 1.4 decades
power law behavior on  the range $80<f<2000\,$Hz.
However,  the Fourier
analysis requires high order harmonic components to \blue{represent the ramp-cliff structures}. It leads \red{to} an
artificial energy \red{transfer} from
low \red{frequencies} (large \red{scales}) to
high \red{frequencies} (small \red{scales}) in Fourier space, causing a less \red{steep} spectrum
\cite{Huang1998,Huang2009PHD}.
 Since both EMD and Hilbert spectral analysis have \red{a} very
local ability, the effect of ramp-cliff \red{structures} is  constrained. }

Due to the presence of ramp-cliff structures, the structure
function analysis fails (figure not shown here, see Ref. \cite{Huang2009PHD}).
However, the Hilbert-based methodology shows a clear inertial range  also for
other moment orders,
up to $q=8$ (not
shown here). Figure \ref{fig:scaling} shows the scaling exponents provided
by Hilbert-based approach $\xi_{\theta}(q)-1$ ($\ocircle$). For comparison,  the scaling exponents
directly estimated by structure functions $\zeta_{\theta}(q)$ ($\diamondsuit$), the scaling exponents
$\zeta_{\theta}(q)$ ($\square$) compiled by \citet{Schmitt2005} for passive
scalar, and the ESS
scaling exponents $\zeta(q)$ (dashed line) for velocity \cite{Arneodo1996}.
 Due to the effect of ramp-cliff structures, the scaling exponents provided
directly by the structure functions \red{seem}
to  saturate when $q>2$.
The scaling exponents $\xi_{\theta}(q)-1$ provided by the Hilbert-based methodology are quite close to the ESS for the longitudinal velocity \cite{Arneodo1996}, indicating a less intermittent scalar field than what
was believed before. We must underline here that the Hilbert-based approach provided the same
exponents as the structure function for the velocity field \cite{Huang2008EPL} when there is no large
scale energetic forcing. The difference found here for the passive scalar case may thus come from the fact
that temperature fluctuations have a strong large scale contribution.
Apparently the ramp-cliff structure is a large scale \red{of} the order of an integral
scale \cite{Sreenivasan1991}.
The cliff  is sharp, and thus is manifested at the small scales: this
\red{may be interpreted as a}   coupling between the large ramp-cliff structures and the small
scales \cite{Sreenivasan1991}. As we  argued above, the inertial range, if it
exists,  is strongly influenced by the these large scale structures.

\section{Discussion and summary}\label{sec:CL}
In summary, based on an assumption of statistical stationarity, we investigated here an analytic model of the second order structure function. By
introducing a cumulative function,  we have found that the structure
function is strongly influenced by the large scales. The large scale contribution range  is found as \red{being} 1.4 decades \red{wide} and the contribution is about 79\%.
We \red{have}  shown numerically that the single scale influence range down to the
small scale is as large as 2 decades.  The Hilbert-based methodology may constrain the large scale  effect \red{to} 0.3 decade.
We then \red{showed} an analysis from a  passive scalar time series with strong ramp-cliff structures, in which the classical structure functions fail.
Surprisingly, the scaling exponents predicted by Hilbert-based approach are almost the same as the scaling exponents for longitudinal velocity in fully developed turbulence, indicating a less intermittent passive scalar statistics than what
was believed before.

  This should be verified \red{using} more databases, but it may be giving an explanation to the
question open for a long time, of why passive scalars, being passive quantities, are more intermittent than the velocity field. We hope that the result  obtained here can contribute to reconsidering the statistical properties of turbulence with large energetic scale structures.

 \begin{acknowledgments}
This work is sponsored by the National Natural Science Foundation of China under Grant No. 10772110. \blue{Z.~M. is also supported by STCSM under grant No. 08JC1409800.} Y.~H.  \red{was}
financed in part by a Ph.D grant from the French Ministry of
Foreign Affairs and by part from university of Lille 1. \blue{Y.H. also
acknowledges
a post-doctoral financial support from Pr. Hermand, EHL of Universit\'e libre
de Bruxelles (U.L.B),  and Pr. Verbanck, STEP of Universit\'e libre
de Bruxelles (U.L.B) during the preparation of  this manuscript.}
We thank Prof. Meneveau for sharing his experimental
velocity database, which is available for download at
C. Meneveau's web page \cite{Meneveauweb}.
\red{Finally, we thank two anonymous referees for useful comments.}
\end{acknowledgments}

\appendix
\section{Analytical expression of the second order structure functions}

 \red{In this Appendix, we show hot to obtain the analytical expressions
 \eqref{eq:analyticalSF} and \eqref{eq:analyticalP} for the second order
 structure functions and its cumulative function \eqref{eq:csf},
 respectively, for a scaling power law spectrum given by Eq. \eqref{eq:pl}. }
 
\red{We substitute Eq. \eqref{eq:pl} into Eq. \eqref{eq:2sf}
\begin{equation}
 S_2(\ell)=\int_{0}^{\infty} cf^{-\beta}(1-\cos(2\pi f \ell))\upd f
\end{equation}
After a scaling transform \red{$f'=2\pi\ell f$}, we have
\begin{equation}
 S_2(\ell)= (2\pi\ell)^{\beta-1} \int_{0}^{\infty} cf'^{-\beta}(1-\cos(f'))\upd f'
\end{equation}
We rewrite the
integration range from $0$ to $f$
\begin{equation}
 S_2(\ell,f)=(2\pi\ell)^{\beta-1} \int_{0}^{f} c x^{-\beta}
\left(1-\cos(x)\right)\upd x
\end{equation}
By applying integration by parts, we  have
\begin{equation}
\begin{array}{lll}
  S_2(\ell,f)&=  \frac{c(2\pi\ell)^{\beta-1}}{1-\beta} \bigg\{
\underbrace{x^{1-\beta}
(1-\cos(x))}_{A}\vert_{0}^{f} \\
&-  \underbrace{\int_{0}^{f} 
x^{1-\beta}\sin(x)\upd x}_{B} \bigg\}
\end{array}
\end{equation}
where $1<\beta <3$.
It is not difficult to show \red{that} $ \lim_{f\to 0}A = 0$. An analytical expression for
$B$ is
\begin{equation}
 B={}_{1}F_2(3/2-\beta/2,3/2,5/2-\beta/2,-f^2/4)
\end{equation}
in which ${}_1F_2$ is a generalized hypergeometric function
\cite{Abramowitz1970Handbook}.
 In the \red{limit}  $f\to \infty$, we have 
\begin{equation}
 \lim_{f\to \infty}A=0,\, \lim_{f\to
\infty}B=\frac{\sqrt{\pi}\,\Gamma(\frac{3}{2}-\frac{\beta}{2})}{2^{\beta-1}\,
\Gamma(\frac { \beta } { 2 } ) }
\end{equation}
We finally obtain Eq. \eqref{eq:analyticalSF} and \eqref{eq:analyticalP}.}

\red{The analytical expression for $\mathcal{Q}$ 
 can be obtained by the same
procedure, which reads as}
\blue{
\begin{widetext}
\begin{equation}
 \mathcal{Q}(f,\ell)=\frac{1}{b(\beta)}
\bigg\{(3-\beta)(\cos(f)-1)\cos(f)f^{1-\beta}+\, h(f,\beta)\,f^{3-\beta}
\bigg\}
\times 100\%\label{eq:analyticalQ}
\end{equation}
\end{widetext}
in which
$b(\beta)=-\sqrt{\pi}\,(3-\beta)\,(2^{1-\beta}-1/2)\,
\Gamma(3/2-\beta/2)\Gamma(\beta/2)^{ -1}$, and
$g(f,\beta)=2\,{}_1F_2\big({3}/{2} $ {} $ - {\beta}/{2},
 {3}/{2},{5}/{2}-{\beta}/{2}, -f^2\big) 
-{}_1F_2\!\left({3}/{2}-{\beta}/{2}, {3}/{2},
 {5}/{2}-{\beta}/{2}, -{f^2}/{4}\right)$, and ${}_{1}F_2$ is again a generalized
hypergeometric function. It is also independent of $\ell$.}
\bibliographystyle{apsrev}

\begin{thebibliography}{42}
\expandafter\ifx\csname natexlab\endcsname\relax\def\natexlab#1{#1}\fi
\expandafter\ifx\csname bibnamefont\endcsname\relax
  \def\bibnamefont#1{#1}\fi
\expandafter\ifx\csname bibfnamefont\endcsname\relax
  \def\bibfnamefont#1{#1}\fi
\expandafter\ifx\csname citenamefont\endcsname\relax
  \def\citenamefont#1{#1}\fi
\expandafter\ifx\csname url\endcsname\relax
  \def\url#1{\texttt{#1}}\fi
\expandafter\ifx\csname urlprefix\endcsname\relax\def\urlprefix{URL }\fi
\providecommand{\bibinfo}[2]{#2}
\providecommand{\eprint}[2][]{\url{#2}}

\bibitem[{\citenamefont{Kolmogorov}(1941)}]{Kolmogorov1941}
\bibinfo{author}{\bibfnamefont{A.~N.} \bibnamefont{Kolmogorov}},
  \bibinfo{journal}{Dokl. Akad. Nauk SSSR} \textbf{\bibinfo{volume}{30}},
  \bibinfo{pages}{299} (\bibinfo{year}{1941}).

\bibitem[{\citenamefont{Frisch}(1995)}]{Frisch1995}
\bibinfo{author}{\bibfnamefont{U.}~\bibnamefont{Frisch}},
  \emph{\bibinfo{title}{{Turbulence: the legacy of AN Kolmogorov}}}
  (\bibinfo{publisher}{Cambridge University Press}, \bibinfo{year}{1995}).

\bibitem[{\citenamefont{Anselmet et~al.}(1984)\citenamefont{Anselmet, Gagne,
  Hopfinger, and Antonia}}]{Anselmet1984}
\bibinfo{author}{\bibfnamefont{F.}~\bibnamefont{Anselmet}},
  \bibinfo{author}{\bibfnamefont{Y.}~\bibnamefont{Gagne}},
  \bibinfo{author}{\bibfnamefont{E.~J.} \bibnamefont{Hopfinger}},
  \bibnamefont{and} \bibinfo{author}{\bibfnamefont{R.~A.}
  \bibnamefont{Antonia}}, \bibinfo{journal}{J. Fluid Mech.}
  \textbf{\bibinfo{volume}{140}}, \bibinfo{pages}{63} (\bibinfo{year}{1984}).

\bibitem[{\citenamefont{Lepore and Mydlarski}(2009)}]{Lepore2009}
\bibinfo{author}{\bibfnamefont{L.}~\bibnamefont{Lepore}} \bibnamefont{and}
  \bibinfo{author}{\bibfnamefont{L.}~\bibnamefont{Mydlarski}},
  \bibinfo{journal}{Phy. Rev. Lett.} \textbf{\bibinfo{volume}{103}},
  \bibinfo{pages}{034501} (\bibinfo{year}{2009}).

\bibitem[{\citenamefont{Sreenivasan and Antonia}(1997)}]{Sreenivasan1997}
\bibinfo{author}{\bibfnamefont{K.}~\bibnamefont{Sreenivasan}} \bibnamefont{and}
  \bibinfo{author}{\bibfnamefont{R.}~\bibnamefont{Antonia}},
  \bibinfo{journal}{Annu. Rev. Fluid Mech.} \textbf{\bibinfo{volume}{29}},
  \bibinfo{pages}{435} (\bibinfo{year}{1997}).

\bibitem[{\citenamefont{Lohse and Xia}(2010)}]{Lohse2010}
\bibinfo{author}{\bibfnamefont{D.}~\bibnamefont{Lohse}} \bibnamefont{and}
  \bibinfo{author}{\bibfnamefont{K.-Q.} \bibnamefont{Xia}},
  \bibinfo{journal}{Ann. Rev. Fluid Mech.} \textbf{\bibinfo{volume}{42}},
  \bibinfo{pages}{335} (\bibinfo{year}{2010}).

\bibitem[{\citenamefont{Bacry et~al.}(1993)\citenamefont{Bacry, Muzy, and
  Arneodo}}]{Bacry1993}
\bibinfo{author}{\bibfnamefont{E.}~\bibnamefont{Bacry}},
  \bibinfo{author}{\bibfnamefont{J.}~\bibnamefont{Muzy}}, \bibnamefont{and}
  \bibinfo{author}{\bibfnamefont{A.}~\bibnamefont{Arneodo}},
  \bibinfo{journal}{J. Statist. Phys.} \textbf{\bibinfo{volume}{70}},
  \bibinfo{pages}{635} (\bibinfo{year}{1993}).

\bibitem[{\citenamefont{Nichols~Pagel et~al.}(2008)\citenamefont{Nichols~Pagel,
  Percival, Reinhall, and Riley}}]{Nicholspagel2008}
\bibinfo{author}{\bibfnamefont{G.}~\bibnamefont{Nichols~Pagel}},
  \bibinfo{author}{\bibfnamefont{D.}~\bibnamefont{Percival}},
  \bibinfo{author}{\bibfnamefont{P.}~\bibnamefont{Reinhall}}, \bibnamefont{and}
  \bibinfo{author}{\bibfnamefont{J.}~\bibnamefont{Riley}},
  \bibinfo{journal}{Physica D} \textbf{\bibinfo{volume}{237}},
  \bibinfo{pages}{665} (\bibinfo{year}{2008}).

\bibitem[{\citenamefont{Percival and Walden}(1993)}]{Percival1993}
\bibinfo{author}{\bibfnamefont{D.}~\bibnamefont{Percival}} \bibnamefont{and}
  \bibinfo{author}{\bibfnamefont{A.}~\bibnamefont{Walden}},
  \emph{\bibinfo{title}{{Spectral Analysis for Physical Applications:
  Multitaper and Conventional Univariate Techniques}}}
  (\bibinfo{publisher}{Cambridge University Press}, \bibinfo{year}{1993}).

\bibitem[{\citenamefont{Huang et~al.}(2008{\natexlab{a}})\citenamefont{Huang,
  Schmitt, Lu, and Liu}}]{Huang2008EPL}
\bibinfo{author}{\bibfnamefont{Y.}~\bibnamefont{Huang}},
  \bibinfo{author}{\bibfnamefont{F.~G.} \bibnamefont{Schmitt}},
  \bibinfo{author}{\bibfnamefont{Z.}~\bibnamefont{Lu}}, \bibnamefont{and}
  \bibinfo{author}{\bibfnamefont{Y.}~\bibnamefont{Liu}},
  \bibinfo{journal}{Europhys. Lett.} \textbf{\bibinfo{volume}{84}},
  \bibinfo{pages}{40010} (\bibinfo{year}{2008}{\natexlab{a}}).

\bibitem[{\citenamefont{Huang et~al.}(2008{\natexlab{b}})\citenamefont{Huang,
  Schmitt, Lu, and Liu}}]{Huang2008TSI}
\bibinfo{author}{\bibfnamefont{Y.}~\bibnamefont{Huang}},
  \bibinfo{author}{\bibfnamefont{F.~G.} \bibnamefont{Schmitt}},
  \bibinfo{author}{\bibfnamefont{Z.}~\bibnamefont{Lu}}, \bibnamefont{and}
  \bibinfo{author}{\bibfnamefont{Y.}~\bibnamefont{Liu}},
  \bibinfo{journal}{Traitement du Signal} \textbf{\bibinfo{volume}{25}},
  \bibinfo{pages}{481} (\bibinfo{year}{2008}{\natexlab{b}}).

\bibitem[{\citenamefont{Huang}(2009
  http://tel.archives-ouvertes.fr/tel-00439605/fr)}]{Huang2009PHD}
\bibinfo{author}{\bibfnamefont{Y.}~\bibnamefont{Huang}}, Ph.D. thesis,
  \bibinfo{school}{Universit\'e des Sciences et Technologies de Lille - Lille
  1, France \& Shanghai University, China} (\bibinfo{year}{2009})
  http://tel.archives-ouvertes.fr/tel-00439605/fr.

\bibitem[{\citenamefont{Sreenivasan}(1991)}]{Sreenivasan1991}
\bibinfo{author}{\bibfnamefont{K.}~\bibnamefont{Sreenivasan}},
  \bibinfo{journal}{Proc. R. Soc. Lond. A} \textbf{\bibinfo{volume}{434}},
  \bibinfo{pages}{165} (\bibinfo{year}{1991}).

\bibitem[{\citenamefont{Shraiman and Siggia}(2000)}]{Shraiman2000}
\bibinfo{author}{\bibfnamefont{B.}~\bibnamefont{Shraiman}} \bibnamefont{and}
  \bibinfo{author}{\bibfnamefont{E.}~\bibnamefont{Siggia}},
  \bibinfo{journal}{Nature} \textbf{\bibinfo{volume}{405}},
  \bibinfo{pages}{639} (\bibinfo{year}{2000}).

\bibitem[{\citenamefont{Warhaft}(2000)}]{Warhaft2000}
\bibinfo{author}{\bibfnamefont{Z.}~\bibnamefont{Warhaft}},
  \bibinfo{journal}{Annu. Rev. Fluid Mech.} \textbf{\bibinfo{volume}{32}},
  \bibinfo{pages}{203} (\bibinfo{year}{2000}).

\bibitem[{\citenamefont{Huang et~al.}(1998)\citenamefont{Huang, Shen, Long, Wu,
  Shih, Zheng, Yen, Tung, and Liu}}]{Huang1998}
\bibinfo{author}{\bibfnamefont{N.~E.} \bibnamefont{Huang}},
  \bibinfo{author}{\bibfnamefont{Z.}~\bibnamefont{Shen}},
  \bibinfo{author}{\bibfnamefont{S.~R.} \bibnamefont{Long}},
  \bibinfo{author}{\bibfnamefont{M.~C.} \bibnamefont{Wu}},
  \bibinfo{author}{\bibfnamefont{H.~H.} \bibnamefont{Shih}},
  \bibinfo{author}{\bibfnamefont{Q.}~\bibnamefont{Zheng}},
  \bibinfo{author}{\bibfnamefont{N.}~\bibnamefont{Yen}},
  \bibinfo{author}{\bibfnamefont{C.~C.} \bibnamefont{Tung}}, \bibnamefont{and}
  \bibinfo{author}{\bibfnamefont{H.~H.} \bibnamefont{Liu}},
  \bibinfo{journal}{Proc. R. Soc. London, Ser. A}
  \textbf{\bibinfo{volume}{454}}, \bibinfo{pages}{903} (\bibinfo{year}{1998}).

\bibitem[{\citenamefont{Huang et~al.}(1999)\citenamefont{Huang, Shen, and
  Long}}]{Huang1999}
\bibinfo{author}{\bibfnamefont{N.~E.} \bibnamefont{Huang}},
  \bibinfo{author}{\bibfnamefont{Z.}~\bibnamefont{Shen}}, \bibnamefont{and}
  \bibinfo{author}{\bibfnamefont{S.~R.} \bibnamefont{Long}},
  \bibinfo{journal}{Annu. Rev. Fluid Mech.} \textbf{\bibinfo{volume}{31}},
  \bibinfo{pages}{417} (\bibinfo{year}{1999}).

\bibitem[{\citenamefont{Rilling et~al.}(2003)\citenamefont{Rilling, Flandrin,
  and Gon\c{c}alv\`es}}]{Rilling2003}
\bibinfo{author}{\bibfnamefont{G.}~\bibnamefont{Rilling}},
  \bibinfo{author}{\bibfnamefont{P.}~\bibnamefont{Flandrin}}, \bibnamefont{and}
  \bibinfo{author}{\bibfnamefont{P.}~\bibnamefont{Gon\c{c}alv\`es}},
  \bibinfo{journal}{IEEE-EURASIP Workshop on Nonlinear Signal and Image
  Processing}  (\bibinfo{year}{2003}).

\bibitem[{\citenamefont{Flandrin et~al.}(2004)\citenamefont{Flandrin, Rilling,
  and Gon{\c{c}}alv\`es}}]{Flandrin2004}
\bibinfo{author}{\bibfnamefont{P.}~\bibnamefont{Flandrin}},
  \bibinfo{author}{\bibfnamefont{G.}~\bibnamefont{Rilling}}, \bibnamefont{and}
  \bibinfo{author}{\bibfnamefont{P.}~\bibnamefont{Gon{\c{c}}alv\`es}},
  \bibinfo{journal}{IEEE Sig. Proc. Lett.} \textbf{\bibinfo{volume}{11}},
  \bibinfo{pages}{112} (\bibinfo{year}{2004}).

\bibitem[{\citenamefont{Cohen}(1995)}]{Cohen1995}
\bibinfo{author}{\bibfnamefont{L.}~\bibnamefont{Cohen}},
  \emph{\bibinfo{title}{{Time-frequency analysis}}}
  (\bibinfo{publisher}{Prentice Hall PTR Englewood Cliffs, NJ},
  \bibinfo{year}{1995}).

\bibitem[{\citenamefont{Flandrin}(1998)}]{Flandrin1998}
\bibinfo{author}{\bibfnamefont{P.}~\bibnamefont{Flandrin}},
  \emph{\bibinfo{title}{{Time-frequency/time-scale analysis}}}
  (\bibinfo{publisher}{Academic Press}, \bibinfo{year}{1998}).

\bibitem[{\citenamefont{Huang}(2005)}]{Huang2005a}
\bibinfo{author}{\bibfnamefont{N.~E.} \bibnamefont{Huang}},
  \emph{\bibinfo{title}{Hilbert-Huang Transform and Its Applications}}
  (\bibinfo{publisher}{World Scientific}, \bibinfo{year}{2005}), chap.
  \bibinfo{chapter}{1. Introduction to the Hilbert-Huang transform and its
  related mathematical problems}, pp. \bibinfo{pages}{1--26}.

\bibitem[{\citenamefont{Long et~al.}(1995)\citenamefont{Long, Huang, Tung, Wu,
  Lin, Mollo-Christensen, and Yuan}}]{Long1995}
\bibinfo{author}{\bibfnamefont{S.~R.} \bibnamefont{Long}},
  \bibinfo{author}{\bibfnamefont{N.~E.} \bibnamefont{Huang}},
  \bibinfo{author}{\bibfnamefont{C.~C.} \bibnamefont{Tung}},
  \bibinfo{author}{\bibfnamefont{M.~L.} \bibnamefont{Wu}},
  \bibinfo{author}{\bibfnamefont{R.~Q.} \bibnamefont{Lin}},
  \bibinfo{author}{\bibfnamefont{E.}~\bibnamefont{Mollo-Christensen}},
  \bibnamefont{and} \bibinfo{author}{\bibfnamefont{Y.}~\bibnamefont{Yuan}},
  \bibinfo{journal}{IEEE Geoscience and Remote Sensing Soc. Lett.}
  \textbf{\bibinfo{volume}{3}}, \bibinfo{pages}{6} (\bibinfo{year}{1995}).

\bibitem[{\citenamefont{Huang et~al.}(2009{\natexlab{a}})\citenamefont{Huang,
  Schmitt, Lu, and Liu}}]{Huang2009Hydrol}
\bibinfo{author}{\bibfnamefont{Y.}~\bibnamefont{Huang}},
  \bibinfo{author}{\bibfnamefont{F.~G.} \bibnamefont{Schmitt}},
  \bibinfo{author}{\bibfnamefont{Z.}~\bibnamefont{Lu}}, \bibnamefont{and}
  \bibinfo{author}{\bibfnamefont{Y.}~\bibnamefont{Liu}}, \bibinfo{journal}{J.
  Hydrol.} \textbf{\bibinfo{volume}{373}}, \bibinfo{pages}{103}
  (\bibinfo{year}{2009}{\natexlab{a}}).

\bibitem[{\citenamefont{Schmitt et~al.}(2009)\citenamefont{Schmitt, Huang, Lu,
  Y., and Fernandez}}]{Schmitt2009JM}
\bibinfo{author}{\bibfnamefont{F.~G.} \bibnamefont{Schmitt}},
  \bibinfo{author}{\bibfnamefont{Y.}~\bibnamefont{Huang}},
  \bibinfo{author}{\bibfnamefont{Z.}~\bibnamefont{Lu}},
  \bibinfo{author}{\bibfnamefont{L.}~\bibnamefont{Y.}}, \bibnamefont{and}
  \bibinfo{author}{\bibfnamefont{N.}~\bibnamefont{Fernandez}},
  \bibinfo{journal}{J. Mar. Sys.} \textbf{\bibinfo{volume}{77}},
  \bibinfo{pages}{473} (\bibinfo{year}{2009}).

\bibitem[{\citenamefont{Rilling and Flandrin}(2006)}]{Rilling2006}
\bibinfo{author}{\bibfnamefont{G.}~\bibnamefont{Rilling}} \bibnamefont{and}
  \bibinfo{author}{\bibfnamefont{P.}~\bibnamefont{Flandrin}},
  \bibinfo{journal}{IEEE International Conference on Acoustics, Speech and
  Signal Processing, 2006. ICASSP 2006 Proceedings. 2006}
  \textbf{\bibinfo{volume}{3}}, \bibinfo{pages}{444} (\bibinfo{year}{2006}).

\bibitem[{\citenamefont{Rilling and Flandrin}(2008)}]{Rilling2008}
\bibinfo{author}{\bibfnamefont{G.}~\bibnamefont{Rilling}} \bibnamefont{and}
  \bibinfo{author}{\bibfnamefont{P.}~\bibnamefont{Flandrin}},
  \bibinfo{journal}{IEEE Trans. Signal Process}  (\bibinfo{year}{2008}).

\bibitem[{\citenamefont{Rilling and Flandrin}(2009)}]{Rilling2009}
\bibinfo{author}{\bibfnamefont{G.}~\bibnamefont{Rilling}} \bibnamefont{and}
  \bibinfo{author}{\bibfnamefont{P.}~\bibnamefont{Flandrin}},
  \bibinfo{journal}{Adv. Adapt. Data Anal.} \textbf{\bibinfo{volume}{1}},
  \bibinfo{pages}{43} (\bibinfo{year}{2009}).

\bibitem[{\citenamefont{Monin and Yaglom}(1971)}]{Monin1971}
\bibinfo{author}{\bibfnamefont{A.~S.} \bibnamefont{Monin}} \bibnamefont{and}
  \bibinfo{author}{\bibfnamefont{A.~M.} \bibnamefont{Yaglom}},
  \emph{\bibinfo{title}{{Statistical fluid mechanics vd II}}}
  (\bibinfo{publisher}{MIT Press Cambridge, Mass}, \bibinfo{year}{1971}).

\bibitem[{\citenamefont{Schmitt et~al.}(1995)\citenamefont{Schmitt, Lovejoy,
  and Schertzer}}]{Schmitt1995GRL}
\bibinfo{author}{\bibfnamefont{F.~G.} \bibnamefont{Schmitt}},
  \bibinfo{author}{\bibfnamefont{S.}~\bibnamefont{Lovejoy}}, \bibnamefont{and}
  \bibinfo{author}{\bibfnamefont{D.}~\bibnamefont{Schertzer}},
  \bibinfo{journal}{Geophys. Res. Lett.} \textbf{\bibinfo{volume}{22}},
  \bibinfo{pages}{1689} (\bibinfo{year}{1995}).

\bibitem[{\citenamefont{Schmitt et~al.}(1999)\citenamefont{Schmitt, Schertzer,
  and Lovejoy}}]{Schmitt1999}
\bibinfo{author}{\bibfnamefont{F.~G.} \bibnamefont{Schmitt}},
  \bibinfo{author}{\bibfnamefont{D.}~\bibnamefont{Schertzer}},
  \bibnamefont{and} \bibinfo{author}{\bibfnamefont{S.}~\bibnamefont{Lovejoy}},
  \bibinfo{journal}{Appl. Stoch. Models and Data Anal.}
  \textbf{\bibinfo{volume}{15}}, \bibinfo{pages}{29} (\bibinfo{year}{1999}).

\bibitem[{\citenamefont{Lohse and M{\"u}ller-Groeling}(1995)}]{Lohse1995PRL}
\bibinfo{author}{\bibfnamefont{D.}~\bibnamefont{Lohse}} \bibnamefont{and}
  \bibinfo{author}{\bibfnamefont{A.}~\bibnamefont{M{\"u}ller-Groeling}},
  \bibinfo{journal}{Phys. Rev. Lett.} \textbf{\bibinfo{volume}{74}},
  \bibinfo{pages}{1747s} (\bibinfo{year}{1995}).

\bibitem[{\citenamefont{Lohse and M{\"u}ller-Groeling}(1996)}]{Lohse1996PRE}
\bibinfo{author}{\bibfnamefont{D.}~\bibnamefont{Lohse}} \bibnamefont{and}
  \bibinfo{author}{\bibfnamefont{A.}~\bibnamefont{M{\"u}ller-Groeling}},
  \bibinfo{journal}{Phys. Rev. E} \textbf{\bibinfo{volume}{54}},
  \bibinfo{pages}{395} (\bibinfo{year}{1996}).

\bibitem[{\citenamefont{Abramowitz and Stegun}(1970)}]{Abramowitz1970Handbook}
\bibinfo{author}{\bibfnamefont{M.}~\bibnamefont{Abramowitz}} \bibnamefont{and}
  \bibinfo{author}{\bibfnamefont{I.~A.} \bibnamefont{Stegun}},
  \emph{\bibinfo{title}{{Handbook of Mathematical Functions}}}
  (\bibinfo{publisher}{Dover, New York}, \bibinfo{year}{1970}).

\bibitem[{\citenamefont{Kang et~al.}(2003)\citenamefont{Kang, Chester, and
  Meneveau}}]{kang2003dta}
\bibinfo{author}{\bibfnamefont{H.}~\bibnamefont{Kang}},
  \bibinfo{author}{\bibfnamefont{S.}~\bibnamefont{Chester}}, \bibnamefont{and}
  \bibinfo{author}{\bibfnamefont{C.}~\bibnamefont{Meneveau}},
  \bibinfo{journal}{J. Fluid Mech.} \textbf{\bibinfo{volume}{480}},
  \bibinfo{pages}{129} (\bibinfo{year}{2003}).

\bibitem[{\citenamefont{Hou et~al.}(1998)\citenamefont{Hou, Wu, Chen, and
  Zhou}}]{Hou1998}
\bibinfo{author}{\bibfnamefont{T.}~\bibnamefont{Hou}},
  \bibinfo{author}{\bibfnamefont{X.}~\bibnamefont{Wu}},
  \bibinfo{author}{\bibfnamefont{S.}~\bibnamefont{Chen}}, \bibnamefont{and}
  \bibinfo{author}{\bibfnamefont{Y.}~\bibnamefont{Zhou}},
  \bibinfo{journal}{Phys. Rev. E} \textbf{\bibinfo{volume}{58}},
  \bibinfo{pages}{5841} (\bibinfo{year}{1998}).

\bibitem[{\citenamefont{Nelkin}(1994)}]{Nelkin1994}
\bibinfo{author}{\bibfnamefont{M.}~\bibnamefont{Nelkin}},
  \bibinfo{journal}{Adv. Phys.} \textbf{\bibinfo{volume}{43}},
  \bibinfo{pages}{143} (\bibinfo{year}{1994}).

\bibitem[{\citenamefont{Huang et~al.}(2009{\natexlab{b}})\citenamefont{Huang,
  Schmitt, Lu, and Liu}}]{Huang2009EPL}
\bibinfo{author}{\bibfnamefont{Y.}~\bibnamefont{Huang}},
  \bibinfo{author}{\bibfnamefont{F.~G.} \bibnamefont{Schmitt}},
  \bibinfo{author}{\bibfnamefont{Z.}~\bibnamefont{Lu}}, \bibnamefont{and}
  \bibinfo{author}{\bibfnamefont{Y.}~\bibnamefont{Liu}},
  \bibinfo{journal}{Europhys. Lett.} \textbf{\bibinfo{volume}{86}},
  \bibinfo{pages}{40010} (\bibinfo{year}{2009}{\natexlab{b}}).

\bibitem[{\citenamefont{Wu and Huang}(2004)}]{Wu2004a}
\bibinfo{author}{\bibfnamefont{Z.}~\bibnamefont{Wu}} \bibnamefont{and}
  \bibinfo{author}{\bibfnamefont{N.~E.} \bibnamefont{Huang}},
  \bibinfo{journal}{Proc. R. Soc. London, Ser. A}
  \textbf{\bibinfo{volume}{460}}, \bibinfo{pages}{1597} (\bibinfo{year}{2004}).

\bibitem[{\citenamefont{Schmitt}(2005)}]{Schmitt2005}
\bibinfo{author}{\bibfnamefont{F.~G.} \bibnamefont{Schmitt}},
  \bibinfo{journal}{Eur. Phys. J. B} \textbf{\bibinfo{volume}{48}},
  \bibinfo{pages}{129} (\bibinfo{year}{2005}).

\bibitem[{\citenamefont{Arneodo et~al.}(1996)\citenamefont{Arneodo, Baudet,
  Belin, Benzi, Castaing, Chabaud, Chavarria, Ciliberto, Camussi, and
  Chilla}}]{Arneodo1996}
\bibinfo{author}{\bibfnamefont{A.}~\bibnamefont{Arneodo}},
  \bibinfo{author}{\bibfnamefont{C.}~\bibnamefont{Baudet}},
  \bibinfo{author}{\bibfnamefont{F.}~\bibnamefont{Belin}},
  \bibinfo{author}{\bibfnamefont{R.}~\bibnamefont{Benzi}},
  \bibinfo{author}{\bibfnamefont{B.}~\bibnamefont{Castaing}},
  \bibinfo{author}{\bibfnamefont{B.}~\bibnamefont{Chabaud}},
  \bibinfo{author}{\bibfnamefont{R.}~\bibnamefont{Chavarria}},
  \bibinfo{author}{\bibfnamefont{S.}~\bibnamefont{Ciliberto}},
  \bibinfo{author}{\bibfnamefont{R.}~\bibnamefont{Camussi}}, \bibnamefont{and}
  \bibinfo{author}{\bibfnamefont{F.}~\bibnamefont{Chilla}},
  \bibinfo{journal}{Europhys. Lett.} \textbf{\bibinfo{volume}{34}},
  \bibinfo{pages}{411} (\bibinfo{year}{1996}).

\bibitem[{\citenamefont{Celani et~al.}(2000)\citenamefont{Celani, Lanotte,
  Mazzino, and M.}}]{Celani2000}
\bibinfo{author}{\bibfnamefont{A.}~\bibnamefont{Celani}},
  \bibinfo{author}{\bibfnamefont{A.}~\bibnamefont{Lanotte}},
  \bibinfo{author}{\bibfnamefont{A.}~\bibnamefont{Mazzino}}, \bibnamefont{and}
  \bibinfo{author}{\bibfnamefont{V.}~\bibnamefont{M.}}, \bibinfo{journal}{Phys.
  Rev. Lett.} \textbf{\bibinfo{volume}{84}}, \bibinfo{pages}{2385}
  (\bibinfo{year}{2000}).

\bibitem[{\citenamefont{Celani et~al.}(2000)\citenamefont{Celani, Lanotte,
  Mazzino, and M.}}]{Meneveauweb}
http://www.me.jhu.edu/\~{}me\-neveau/datasets.html.
\end{thebibliography}

\end{document}